\documentclass[preprint,twocolumn,showpacs,superscriptaddress,prb,aps,floatfix]{revtex4-2}
\usepackage{graphicx}
\usepackage{rotating}
\usepackage{amsmath}
\usepackage[usenames,dvipsnames]{color}
\usepackage{float}
\usepackage{rotating}
\usepackage{booktabs}
\usepackage{multirow}
\usepackage{longtable}
\usepackage[colorlinks,linkcolor=blue,citecolor=blue]{hyperref}
\usepackage[T1]{fontenc}
\usepackage{bm}
\usepackage{upgreek}
\usepackage{comment}
\makeatletter

\newcommand{\Rmnum}[1]{\expandafter\@slowromancap\romannumeral #1@}

\newcommand{\DC}[1]{\textcolor{red}{To Domenico and Cesare:}}
\makeatletter
\hfuzz=\maxdimen
\usepackage{mhchem}
\usepackage{amsmath}

\begin{document}

	\title{Stacking-dependent thermoelectric transport in layered Sc$_2$Si$_2$Te$_6$ from first principles}
	
	\author{Zhongjuan Han}
	\affiliation{Key Laboratory of Advanced Materials and Devices for Post-Moore Chips, Ministry of Education, University of Science and Technology Beijing, Beijing 100083, China}
	\affiliation{School of Mathematics and Physics, University of Science and Technology Beijing, Beijing 100083, China}

	\author{Wu Xiong}
	\affiliation{Key Laboratory of Advanced Materials and Devices for Post-Moore Chips, Ministry of Education, University of Science and Technology Beijing, Beijing 100083, China}
	\affiliation{School of Mathematics and Physics, University of Science and Technology Beijing, Beijing 100083, China}

	\author{Zhonghao Xia}
	\affiliation{Key Laboratory of Advanced Materials and Devices for Post-Moore Chips, Ministry of Education, University of Science and Technology Beijing, Beijing 100083, China}
	\affiliation{School of Mathematics and Physics, University of Science and Technology Beijing, Beijing 100083, China}

	\author{Weitong Huang}
	\affiliation{School of Mathematics and Physics, University of Science and Technology Beijing, Beijing 100083, China}

	\author{Jiangang He}
	\email{jghe2021@ustb.edu.cn}
	\affiliation{Key Laboratory of Advanced Materials and Devices for Post-Moore Chips, Ministry of Education, University of Science and Technology Beijing, Beijing 100083, China}
	\affiliation{School of Mathematics and Physics, University of Science and Technology Beijing, Beijing 100083, China}
	
	\date{\today}

	\begin{abstract}
	Stacking polymorphism is a common characteristic of van der Waals layered materials and can substantially modify their physical properties. Here, based on first-principles calculations combined with electron and phonon transport theories, we systematically investigate the thermodynamic stability, electronic structure, lattice dynamics, and thermoelectric performance of Sc$_2$Si$_2$Te$_6$ with three high-symmetry stacking sequences, namely, AA, AB, and ABC. We find that the AA- and AB-stacked structures are nearly degenerate in energy with the experimentally reported ABC phase, and that the maximum sliding barrier among these stacking sequences is only about 10~meV/atom, thereby accounting for the stacking faults observed experimentally. These three stacking sequences exhibit distinct electronic structures, with the conduction-band minimum being highly sensitive to the stacking sequence. As a consequence, the conduction-band degeneracies are 12, 2, and 8 for the ABC, AA, and AB stackings, respectively, leading to markedly different electronic transport properties near the band edge. The lattice thermal conductivity is governed primarily by three-phonon scattering, whereas four-phonon scattering provides an additional reduction, particularly in the ABC stacking. Among the three structures, the AB stacking exhibits the lowest lattice thermal conductivity owing to its stronger three-phonon scattering and lower phonon group velocity. As a result, the maximum thermoelectric figure of merit, $ZT$, is achieved in the ABC structure, followed closely by the AB structure, whereas the AA structure shows a substantially reduced value. These results demonstrate that the stacking sequence exerts a non-negligible influence on the thermoelectric performance of Sc$_2$Si$_2$Te$_6$ and suggest that suppressing the formation of the AA stacking is important for achieving high thermoelectric performance.
	\end{abstract}
	
	\maketitle
	\section{Introduction}
	Thermoelectric materials enable the direct interconversion between thermal and electrical energy and therefore hold great promise for waste-heat recovery and solid-state refrigeration, attracting considerable attention in the materials community~\cite{science.1159725,2022Chemical}. Their performance is commonly quantified by the dimensionless figure of merit $ZT = S^2\sigma T/(\kappa_{\mathrm{L}} + \kappa_{\mathrm{e}})$, where $S$, $S^2\sigma$, $T$, $\kappa_{\mathrm{L}}$, and $\kappa_{\mathrm{e}}$ denote the Seebeck coefficient, power factor (PF), absolute temperature, lattice thermal conductivity, and electronic thermal conductivity, respectively. A central challenge in enhancing $ZT$ arises from the strong coupling between $\sigma$ and $S$, which generally exhibit opposite dependences on carrier concentration ($n$) and effective mass. For example, insulators and lightly doped semiconductors typically possess a large $S$ but extremely low $\sigma$ owing to the scarcity of charge carriers~\cite{2008Complex}, whereas heavily doped semiconductors and metals usually exhibit high $\sigma$ but small $S$. In general, a large $S$ requires a high density-of-states effective mass, while high $\sigma$ relies on a small transport effective mass and high carrier mobility. Band engineering has therefore emerged as one of the most effective strategies for improving the PF~\cite{moshwan2019realizing,tan2017improving,long2023band,pei2012thermoelectric}. Since $\kappa_{\mathrm{e}}$ is proportional to $\sigma$, as described by the Wiedemann--Franz law~\cite{solidstatephysics}, suppressing $\kappa_{\mathrm{L}}$ provides another effective route to enhancing $ZT$. According to simple kinetic theory~\cite{tritt2005thermal}, $\kappa_{\mathrm{L}} = \frac{1}{3}C_{\mathrm{v}}\nu_{\mathrm{g}}^2\tau$, where $C_{\mathrm{v}}$, $\nu_{\mathrm{g}}$, and $\tau$ denote the heat capacity, phonon group velocity, and phonon relaxation time, respectively. Therefore, materials with short $\tau$ and low $\nu_{\mathrm{g}}$ generally exhibit low $\kappa_{\mathrm{L}}$. Effective strategies for reducing $\tau$ include point-defect and alloy-disorder engineering~\cite{Mao03042018}, manipulation of nanostructured precipitates~\cite{doi:10.1126/science.1092963,2012High,doi:10.1126/science.1156446}, enhancement of anharmonicity induced by lone-pair electrons~\cite{PhysRevLett.107.235901,2013Lone}, and regulation of rattling modes~\cite{2015Impact,2016Ultralow}. In addition, $\nu_{\mathrm{g}} \propto \sqrt{K/M}$, where $K$ and $M$ represent the chemical bond strength and atomic mass, respectively. Consequently, compounds with soft chemical bonding~\cite{https://doi.org/10.1002/adfm.202108532,https://doi.org/10.1002/advs.202417292} and heavy constituent atoms typically possess low $\kappa_{\mathrm{L}}$.
	
	Layered semiconductors provide a promising platform for balancing $S$, $\sigma$, and $\kappa_{\mathrm{L}}$. To date, many layered thermoelectric materials have been reported~\cite{10.1063/5.0074489}, such as Bi$_2$Te$_3$~\cite{wright1958thermoelectric}, BiCuSeO~\cite{zhao2010bi}, and SnSe~\cite{zhao2014ultralow}. In 2019, Luo et al. reported that Sb$_2$Si$_2$Te$_6$, a honeycomb van der Waals (vdW) compound, exhibits excellent thermoelectric performance, with $\kappa_{\mathrm{L}}$ as low as 0.29~Wm$^{-1}$K$^{-1}$ and a $ZT$ value of 1.65 at 823~K~\cite{LUO}. Subsequently, numerous studies have been carried out on $A_2$Si$_2$Te$_6$ ($A$ = Sb, Bi, and Sc) compounds~\cite{as,jacs,CHEN2022135968}. For example, in 2022, Jang et al. conducted a comparative study of the thermoelectric properties of Sb$_2$Si$_2$Te$_6$ and Bi$_2$Si$_2$Te$_6$~\cite{doi:10.1021/acsami.1c23351}. In 2024, Dou et al. proposed, based on first-principles calculations, replacing Sb atoms in ABC-stacked Sb$_2$Si$_2$Te$_6$ with Sc atoms, leading to a synergistic enhancement of band-valley degeneracy and anisotropy. Consequently, the $ZT$ value increased from 2.76 for Sb$_2$Si$_2$Te$_6$ to 3.51 for Sc$_2$Si$_2$Te$_6$~\cite{Sc2Si2Te6}, demonstrating that Sc$_2$Si$_2$Te$_6$ is a promising thermoelectric material.
	
	On the other hand, weak interlayer coupling gives rise to low sliding barriers between neighboring layers. For example, the interlayer sliding barriers in graphene~\cite{gragap}, MoS$_2$~\cite{MoS2gap}, CrSBr~\cite{CrSBrgap}, and SnP$_2$Se$_6$~\cite{bwx9-xk4d} are approximately 8, 12, 13, and 10~meV/atom, respectively. Therefore, these materials often exhibit multiple stacking sequences or stacking faults~\cite{https://doi.org/10.1002/zaac.202200234}. Experimental studies demonstrate that the vdW layered $A_2$Si$_2$Te$_6$ family is prone to the formation of stacking faults during growth or synthesis~\cite{https://doi.org/10.1002/zaac.202200234}, thereby giving rise to various stacking configurations. A previous study showed that stacking-fault disorder between layers, as revealed by transmission electron microscopy (TEM), is responsible for the low $\kappa_{\mathrm{L}}$ of In$_2$Ge$_2$Te$_6$~\cite{C7TA04810F}. In addition, several studies have demonstrated that altering the stacking mode can significantly regulate the thermoelectric performance of materials. For example, first-principles studies on two stacking structures of Pb$_2$Bi$_2$Te$_5$ indicated that the thermoelectric performance of stacking sequence B is far superior to that of stacking sequence A~\cite{D1CP00270H}. However, current research on Sc$_2$Si$_2$Te$_6$ has primarily focused on the ABC stacking configuration, whereas the possible existence of other stacking configurations and their impact on thermoelectric performance remain unclear.
	
	Therefore, we performed a systematic first-principles study of the stacking sequences of Sc$_2$Si$_2$Te$_6$ and their effects on its thermoelectric properties. Our results show that, in addition to the well-studied ABC stacking structure, the other two high-symmetry stacking sequences, AA and AB, possess nearly identical total energies and small sliding barriers. These three stacking structures exhibit distinct electronic structures, with different conduction-band degeneracies and electrical transport properties at a carrier concentration of $5 \times 10^{19}$~cm$^{-3}$. However, the differences among the three stacking sequences decrease with increasing carrier concentration, leading to comparable electronic transport properties under both $n$-type and $p$-type doping at a carrier concentration of $2 \times 10^{20}$~cm$^{-3}$. The $\kappa_{\mathrm{L}}$ values of these three stacking sequences differ by approximately 10\%, with the AB structure exhibiting the lowest value and the AA stacking sequence the highest. Under $n$-type doping, the highest $ZT$ is consistently achieved along the out-of-plane direction, mainly owing to favorable electronic transport properties, with values of 1.74, 1.72, and 1.33 for AB, ABC, and AA stacking, respectively. Under $p$-type doping, high $ZT$ values are also obtained along the out-of-plane direction, namely 1.59, 1.46, and 1.04 for AB, ABC, and AA stacking, respectively, mainly originating from the ultralow $\kappa_{\mathrm{L}}$ and weak polar optical phonon scattering along this direction. Our results reveal that the AB and ABC stacking sequences of Sc$_2$Si$_2$Te$_6$ exhibit comparable thermoelectric performance, whereas the AA stacking structure shows inferior thermoelectric performance and should therefore be avoided through appropriate synthesis strategies.

	\begin{figure*}[!ht]
		\centering
		\includegraphics[width=1.0\linewidth]{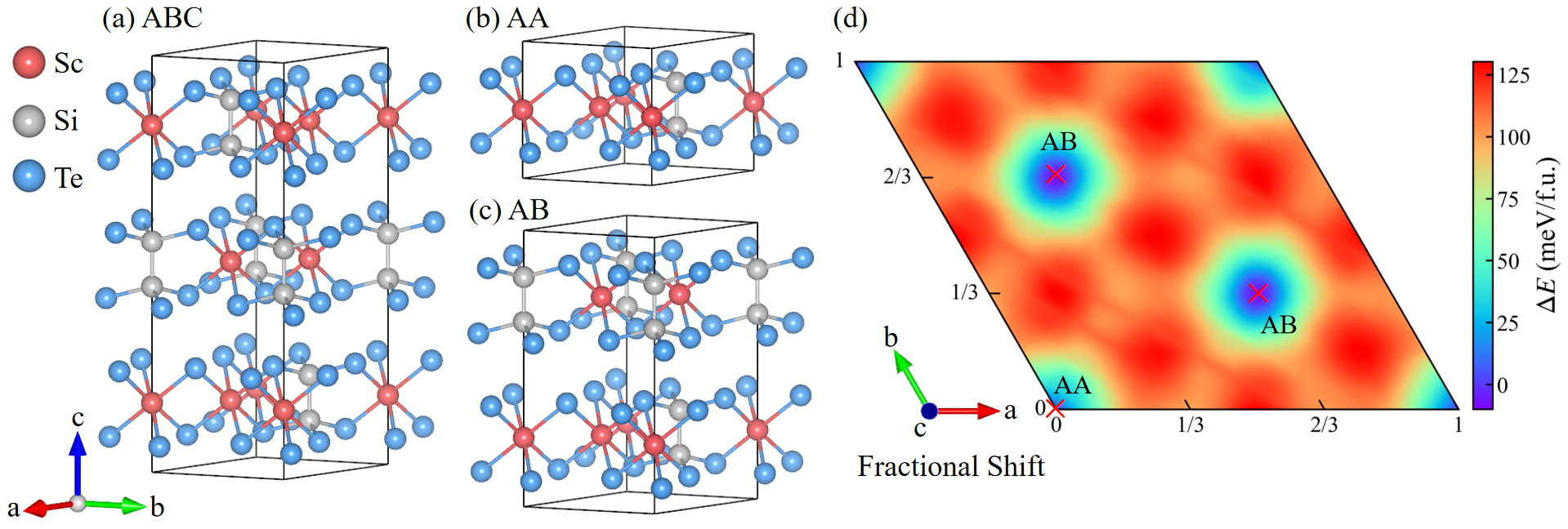}
		\caption{Crystal structures of Sc$_2$Si$_2$Te$_6$ in ABC (a), AA (b), and AB (c) stacking sequences. (d) Potential energy surface obtained by sliding one layer relative to the other in the bilayer stacking sequence within a unit cell.}
		\label{crystal structure}
	\end{figure*}

	\begin{table*}[!ht]
		\scriptsize
		\setlength{\tabcolsep}{6pt} 
		\renewcommand{\arraystretch}{1.3} 
		\centering
		\caption{Comparison of the physical properties of Sc$_2$Si$_2$Te$_6$ with different stacking configurations, including the space group, energy difference relative to ABC [$\Delta E = E - E_{\mathrm{ABC}}$], lattice constants ($a$ and $c$; values in parentheses are obtained using optB86b functional~\cite{PhysRevB.83.195131}), band gap ($E_{\mathrm{g}}^{\mathrm{mBJ}}$), high-frequency dielectric constants ($\epsilon_{\perp}^{\infty}$ and $\epsilon_{\parallel}^{\infty}$), and static dielectric constants ($\epsilon_{\perp}^{0}$ and $\epsilon_{\parallel}^{0}$).} 
		
		\begin{tabular}{@{}cccccccc@{}}
			\toprule
			Types & Space group & $\Delta E$ (meV/f.u.) & $a$ (\AA) & $c$ (\AA) & $E^\mathrm{mBJ}_\mathrm{g}$ (eV) & $\varepsilon^{\infty}_{\perp}$, $\varepsilon^{\infty}_{\parallel}$ & $\varepsilon^0_{\perp}$, $\varepsilon^0_{\parallel}$ \\
			\midrule
			ABC  & $R\overline{3}$   &  0.000 & 6.9537 (7.0174) & 21.1818 (21.1971) & 0.73 & 13.89,8.71 & 19.89,9.05 \\
			AA   & $P\overline{3}$   &  7.433 & 6.9520 (7.0138) &  7.0699 (7.0870)  & 0.66 & 13.92,8.70 & 19.84,9.03 \\
			AB   & $P\overline{3}$   & -0.005 & 6.9547 (7.0142) & 14.1154 (14.1336) & 0.73 & 13.90,8.73 & 19.89,9.08 \\
			\bottomrule
		\end{tabular}
		\label{1111}
	\end{table*}    
	
	\maketitle
	\section{Computational methods}
	All density functional theory (DFT)~\cite{dft1,dft2} calculations were performed using the Vienna \textit{Ab initio} Simulation Package (\textsc{VASP})~\cite{vasp1,vasp2}. The PBEsol functional~\cite{pbesol.PhysRevLett.100.136406} within the generalized gradient approximation (GGA)~\cite{GGA}, projector augmented-wave (PAW) pseudopotentials~\cite{PhysRevB.50.17953,PhysRevB.59.1758}, and a plane-wave basis set with a kinetic-energy cutoff of 520~eV were employed throughout the calculations. For ABC- and AA-type Sc$_2$Si$_2$Te$_6$, which contain 10 atoms in the primitive cell, a $\Gamma$-centered $8 \times 8 \times 8$ $k$-point mesh was used. For the AB phase, which contains 20 atoms in the primitive cell, a $\Gamma$-centered $8 \times 8 \times 4$ $k$-point mesh was adopted. The electronic self-consistency criterion was set to $10^{-8}$~eV, and the projection operators were evaluated in reciprocal space (\texttt{LREAL = FALSE}) to ensure high numerical accuracy.
	
	The electronic structures and band gaps were calculated using the modified Becke--Johnson (mBJ) exchange potential~\cite{mbj}, with spin--orbit coupling (SOC) included. The electronic transport coefficients ($\sigma$, $S$, and $\kappa_{\mathrm{e}}$) were evaluated within the momentum relaxation time approximation (MRTA), as implemented in the AMSET code~\cite{amset}. In these calculations, acoustic deformation-potential (ADP), polar optical phonon (POP), and ionized-impurity (IMP) scattering mechanisms were all taken into account. The deformation potentials required for the AMSET calculations were obtained using the mBJ exchange potential, whereas the elastic constants, dielectric constants, and phonon frequencies at the $\Gamma$ point were computed using the PBEsol functional.
	
	The second-order force constants were calculated using the finite-displacement method, as implemented in the PHONOPY code~\cite{phonopy}, with $3 \times 3 \times 3$ and $3 \times 3 \times 2$ supercells for systems containing 10 and 20 atoms in the primitive cell, respectively. A $2 \times 2 \times 2$ $k$-point mesh was used for the former, whereas only the $\Gamma$ point ($1 \times 1 \times 1$) was used for the latter. The third- and fourth-order force constants were extracted using the compressive sensing lattice dynamics (CSLD) method~\cite{PhysRevLett.113.185501}, as implemented in the ALAMODE package~\cite{alamode,alamode2}. The second-order force constants at finite temperature were renormalized within self-consistent phonon (SCPH) theory~\cite{scph1,scph2,scph3}. The $\kappa_{\mathrm{L}}$, including both three- and four-phonon scattering processes, was obtained by solving the Peierls--Boltzmann transport equation using the FourPhonon package~\cite{HAN2022108179,ShengBTE_2014}. For systems with 10- and 20-atom primitive cells, $q$-point meshes of $12 \times 12 \times 10$ and $12 \times 12 \times 20$ were adopted, respectively; the corresponding convergence tests are shown in Fig.~\textcolor{red}{S1}. The calculation of four-phonon scattering processes was accelerated using the sampling method~\cite{guo2024sampling}. In addition, the off-diagonal terms of the heat-flux operator, corresponding to the coherent phonon contribution, were included within the unified theory of thermal transport~\cite{simoncelli2019unified}.

	\maketitle
	\section{Results and discussion}
	\subsection{Crystal structures.}
	Sc$_2$Si$_2$Te$_6$ adopts a vdW layered structure, in which each single-layer slab consists of ScTe$_6$ octahedra and Te$_6$ octahedra centered by Si--Si dimers (Si$_2$). The alternating arrangement of ScTe$_6$ and Si$_2$Te$_6$ octahedra within the slab gives rise to $C_{3d}$ point-group symmetry. Experimentally, Sc$_2$Si$_2$Te$_6$ crystallizes in an ABC stacking sequence, where the Si$_2$ dimers are aligned with Sc atoms in adjacent layers, corresponding to the space group $R\overline{3}$, as shown in Fig.~\ref{crystal structure}(a). However, other stacking sequences are also possible. In the AA stacking sequence, both the Sc atoms and Si$_2$ dimers are directly aligned with their counterparts in adjacent layers, corresponding to the space group $P\overline{3}$. The AB stacking sequence, which also belongs to the space group $P\overline{3}$, is a bilayer structure obtained by fixing one layer and translating the other by $1/3$ ($2/3$) and $2/3$ ($1/3$) of the lattice vectors $\mathbf{a}$ and $\mathbf{b}$, respectively. We further performed a potential-energy-surface screening for stacking structures containing two single layers within one unit cell. As shown by the mapped potential energy surface (PES) in Fig.~\ref{crystal structure}(d), two high-symmetry structures can be obtained by translating the top layer along the in-plane lattice vectors, accompanied by relaxation of the interlayer distance. The PES further indicates that the AA and AB stacking configurations correspond to two local energy minima, with the AB sequence being slightly lower in energy than the AA sequence. Notably, any two adjacent layers in the ABC stacking adopt the AB stacking sequence, while the top layer is translated relative to the middle layer by $1/3$ ($-1/3$) and $-1/3$ ($1/3$) along the $\mathbf{a}$ and $\mathbf{b}$ directions, respectively.
	
	Table~\ref{1111} presents the energy differences relative to the experimental ABC stacking structure, together with the lattice constants of the three stacking sequences calculated using the PBEsol functional. First, the calculated lattice constants of the ABC structure are in excellent agreement with the experimental values, with relative errors of only 0.6~\% and 0.3~\% for $a$ and $c$, respectively~\cite{https://doi.org/10.1002/zaac.202200234}. These deviations are substantially smaller than those reported in previous theoretical calculations ($\sim$ -2\% and $\sim$ -5\%)~\cite{Sc2Si2Te6}, providing a reliable basis for predicting the thermoelectric properties of these materials. Second, the energy difference between the ABC and AB stacking structures is negligible, and that between the AA and ABC stacking structures is also less than 1~meV/atom. This result indicates that all three stacking sequences possess comparable thermodynamic stability. As shown in Fig.~\ref{crystal structure}(d), the energy barrier for transforming the AA stacking sequence into the AB sequence is approximately 10~meV/atom, comparable to those of SnP$_2$Se$_6$~\cite{bwx9-xk4d} and MoS$_2$~\cite{MoS2gap}. Moreover, the in-plane lattice constants ($a$ and $b$) of the three stacking sequences are nearly identical, whereas the out-of-plane lattice constant ($c$) is approximately proportional to the number of stacked layers, which is a characteristic feature of vdW materials. The interlayer distances in the AB and ABC sequences are identical and slightly smaller than that in the AA sequence, likely owing to the stronger interatomic repulsion that occurs when atoms in adjacent layers are directly aligned.

    \begin{figure*}[!ht]
	\centering
	\includegraphics[width=1.0\linewidth]{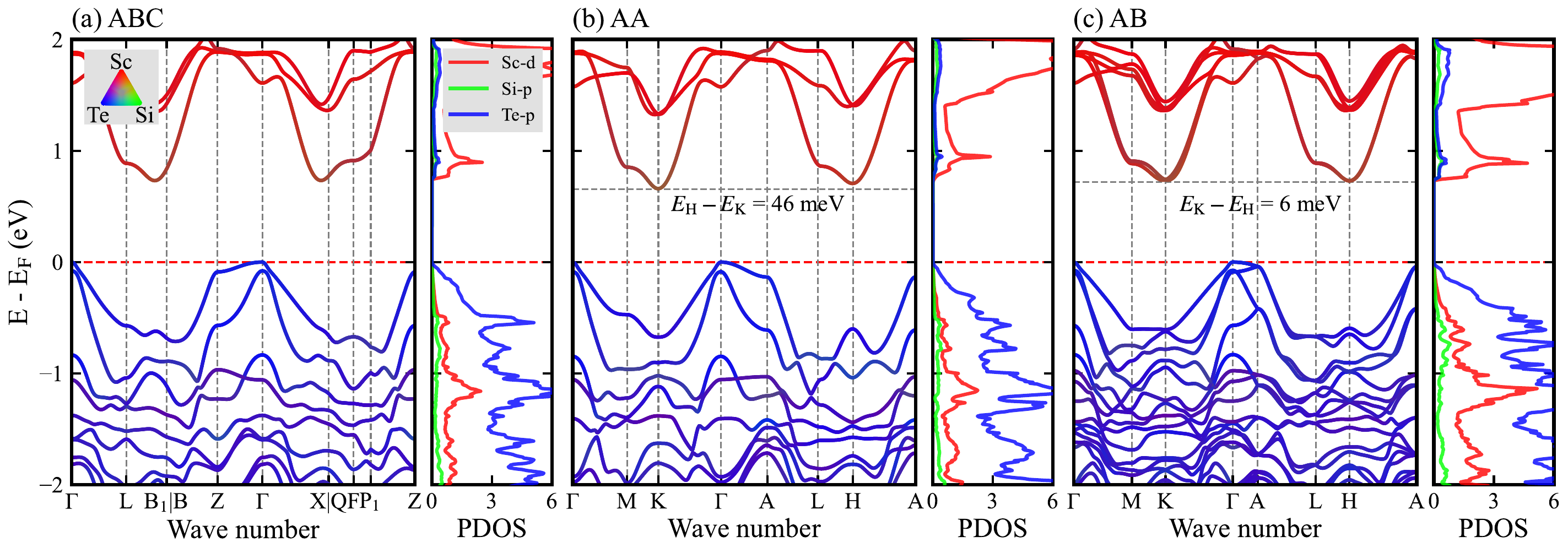}
	\caption{Color-coded electronic band structures and projected density of states (PDOS) of Sc$_2$Si$_2$Te$_6$ for (a) ABC, (b) AA, and (c) AB stacking configurations, calculated with spin-orbit coupling (SOC) included.}
	\label{band}
    \end{figure*}

    \begin{table*}[!ht]
	\scriptsize
	\setlength{\tabcolsep}{4pt} 
	\renewcommand{\arraystretch}{1.3} 
	\centering
	\caption{Maximum power factor (PF$_{\mathrm{max}}$, $\mu$W~cm$^{-1}$~K$^{-2}$) and the corresponding carrier concentration ($n_{\mathrm{h}}/n_{\mathrm{e}}$, cm$^{-3}$), electrical conductivity ($\sigma$, S~m$^{-1}$), Seebeck coefficient ($S$, $\mu$V~K$^{-1}$), and band effective mass ($m_{\mathrm{b}}^{*}$, in units of $m_{0}$) of Sc$_2$Si$_2$Te$_6$ with different stacking configurations under $p$-type and $n$-type doping at 300~K.}
	\begin{tabular}{@{}cccccccccccc@{}}
		\toprule
		& \multirow{2}{*}{Types} & \multicolumn{5}{c}{$p$-type} & \multicolumn{5}{c}{$n$-type} \\  
		\cmidrule(lr){3-7} \cmidrule(lr){8-12}
		& & $n_\mathrm{h}$ & $\sigma$ & $S$ & PF$_\mathrm{max}$ & $m^*_\mathrm{b}$ & $n_\mathrm{e}$ & $\sigma$ & $S$ & PF$_\mathrm{max}$ & $m^*_\mathrm{b}$ \\
		\midrule
		\multirow{3}{*}{in-plane}  & ABC    & 2$\times$10$^{20}$ & 68226  & 183 & 22.8 & 0.17 ($\Gamma$--X) & 1$\times$10$^{21}$ & 348928 & -150 & 78.5 & 0.49 ($\Sigma$--B$_1$) \\
		& AA  & 2$\times$10$^{20}$ & 65709  & 168 & 18.5 & 0.17 ($\Gamma$--M) & 1$\times$10$^{21}$ & 302355 & -152 & 69.9 & 0.39 (K--$\Gamma$) \\
		& AB  & 2$\times$10$^{20}$ & 75725  & 177 & 23.7 & 0.18 ($\Gamma$--M) & 1$\times$10$^{21}$ & 342946 & -149 & 76.1 & 0.38 (K--$\Gamma$) \\
		\hline
		\multirow{3}{*}{out-of-plane}  & ABC    & 4$\times$10$^{20}$ &  52024 & 167 & 14.5 & 3.54 ($\Gamma$--Z) & 3$\times$10$^{20}$ & 175 & -213 & 0.08 & 0.41 ($\Gamma$--Z) \\
		& AA  & 1$\times$10$^{21}$ & 115748 & 121 & 16.9 & 2.29 ($\Gamma$--A) & 1$\times$10$^{20}$ & 897 & -214 & 0.41 & 0.40 ($\Gamma$--A) \\
		& AB  & 3$\times$10$^{20}$ &  48685 & 170 & 14.1 & 3.81 ($\Gamma$--A) & 3$\times$10$^{20}$ & 153 & -181 & 0.05 & 0.40 ($\Gamma$--A) \\
		\bottomrule
	\end{tabular}
	\label{pfmax}
    \end{table*}    

	\subsection{Electronic structures.} 
	Although the energy differences among the three stacking sequences are very small, their electronic structures differ substantially. The electronic band structures and density of states (DOS), calculated using the mBJ functional with SOC included, are shown in Fig.~\ref{band}, and the corresponding indirect band gaps are listed in Table~\ref{1111}. All three structures are identified as indirect-band-gap semiconductors. The valence-band maximum (VBM) is located at the $\Gamma$ point, i.e., the center of the Brillouin zone, with a valley ($k$-point) degeneracy of 1, whereas the location of the conduction-band minimum (CBM) depends sensitively on the stacking sequence. For the ABC stacking configuration, the CBM is located along both the L--B$_1$ and $\Gamma$--X directions. These band extrema correspond to six equivalent $k$ points in the Brillouin zone, giving rise to a valley degeneracy of 12. In contrast, the CBM of the AA stacking structure is located at the high-symmetry K point, with a valley degeneracy of 2. For the AB stacking structure, the CBM appears at H point, which is only 6 meV in energy lower than K point (both them have valley degeneracy of 2). In addition, two degenerate bands occur at K and H points. Therefore, the total conduction-band degeneracy of the AB stacking structure is 8 [$N_{\mathrm{vk}}(\mathrm{K})$ = $N_{\mathrm{vk}}(\mathrm{H})$ = 2, $N_{\mathrm{vo}}(\mathrm{K})$ = 2, $N_{\mathrm{vo}}(\mathrm{H})$ = 2]. The AA stacking configuration exhibits a band gap of approximately 0.66~eV, whereas the AB and ABC configurations have band gaps of approximately 0.73~eV. This trend is similar to that reported for SnP$_2$Se$_6$, where the AA stacking sequence also exhibits the smallest band gap~\cite{bwx9-xk4d}. The DOS indicates that the conduction-band region of all three structures is mainly contributed by the Sc-$d$ orbitals, while the valence-band region is dominated by the Te-$p$ orbitals.
	
	In addition, we calculated the band effective masses ($m^*_\mathrm{b}$) at the VBM and CBM along different $k$ paths for the three stacking configurations using VASPKIT~\cite{vaspkit}; the results are summarized in Table~\ref{pfmax}. Within the single-parabolic-band approximation, a small $m^*_\mathrm{b}$ generally corresponds to high carrier mobility $\mu$ and electrical conductivity $\sigma$, whereas a large $m^*_\mathrm{b}$ is often associated with a large Seebeck coefficient $S$. For the AA stacking structure, the hole effective masses along the $\Gamma$--A direction ($c$ axis) and the $\Gamma$--M direction ($ab$ plane) are 2.29 and 0.17 $m_0$, respectively. Therefore, under hole doping, in-plane $\sigma$ is expected to be higher than that along the out-of-plane, whereas $S$ is expected to show the opposite trend. Notably, the conduction-band energy along the $\Gamma$--A direction ($c$ axis) is relatively high and thus contributes little to electron transport. Consequently, the electron transport performance along the out-of-plane direction is expected to be relatively poor for the AA stacking structure. The out-of-plane $m^*_\mathrm{b}$ values of the AB and ABC structures under hole doping are larger than those of the AA structure, whereas the $m^*_\mathrm{b}$ values for the other doping types and transport directions are nearly identical among the three structures. These results indicate that, although the stacking sequence has only a limited influence on the band dispersion, as reflected by $m^*_\mathrm{b}$, it has a pronounced effect on the band degeneracy. This effect is particularly evident in the AB stacking configuration, where the lattice periodicity along the $c$ axis is doubled and the corresponding Brillouin zone is folded along this direction. As discussed above, band degeneracy plays an important role in determining $S$ and can also affect electron--phonon scattering~\cite{park2021band}, thereby directly influencing $\mu$.

	\begin{figure*}[!ht]
		\centering
		\includegraphics[width=1.0\linewidth]{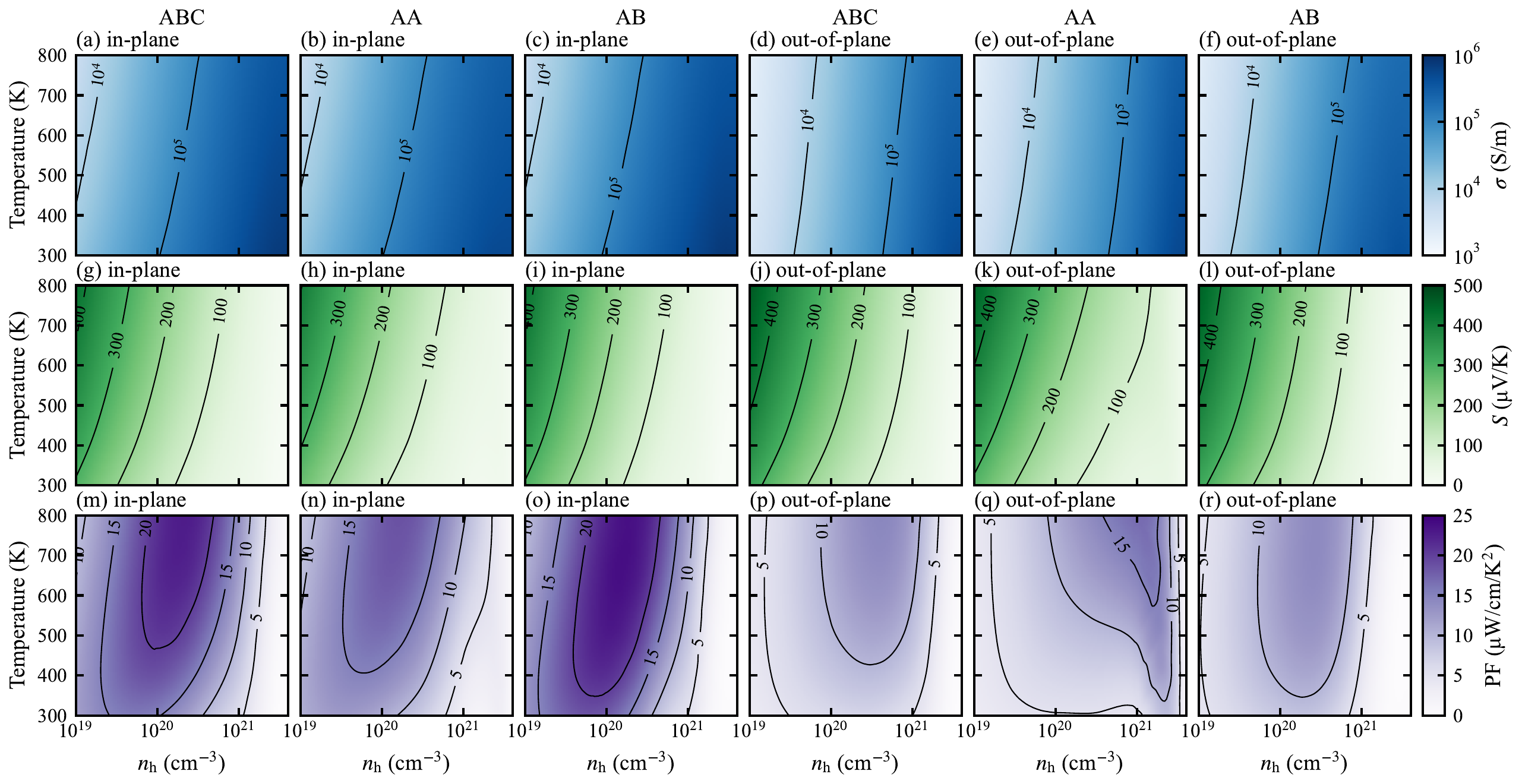}
		\caption{Comparison of electronic transport properties under $p$-type doping for three stacking configurations (ABC, AA, and AB) of Sc$_2$Si$_2$Te$_6$: (a)--(c) in-plane electrical conductivity, (d)--(f) out-of-plane electrical conductivity, (g)--(i) in-plane Seebeck coefficient, (j)--(l) out-of-plane Seebeck coefficient, (m)--(o) in-plane power factor, and (p)--(r) out-of-plane power factor.}
		\label{p}
	\end{figure*}

	\begin{figure*}[!ht]
		\centering
		\includegraphics[width=1.0\linewidth]{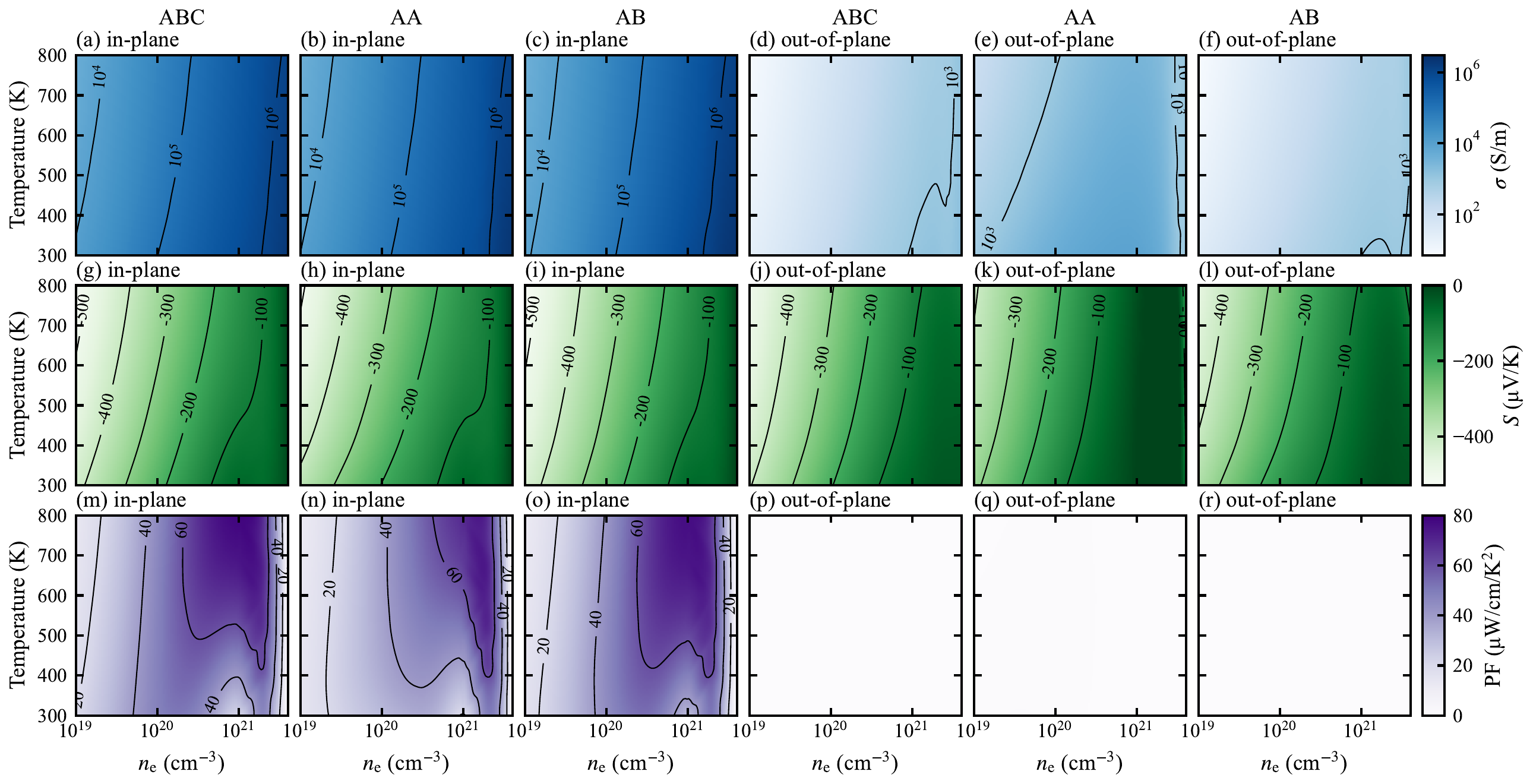}
		\caption{Comparison of electronic transport properties under $n$-type doping for three stacking configurations of Sc$_2$Si$_2$Te$_6$: (a)--(c) in-plane electrical conductivity, (d)--(f) out-of-plane electrical conductivity, (g)--(i) in-plane Seebeck coefficient, (j)--(l) out-of-plane Seebeck coefficient, (m)--(o) in-plane power factor, and (p)--(r) out-of-plane power factor.}
		\label{n}
	\end{figure*}

	\begin{figure*}[!ht]
		\centering
		\includegraphics[width=1.0\linewidth]{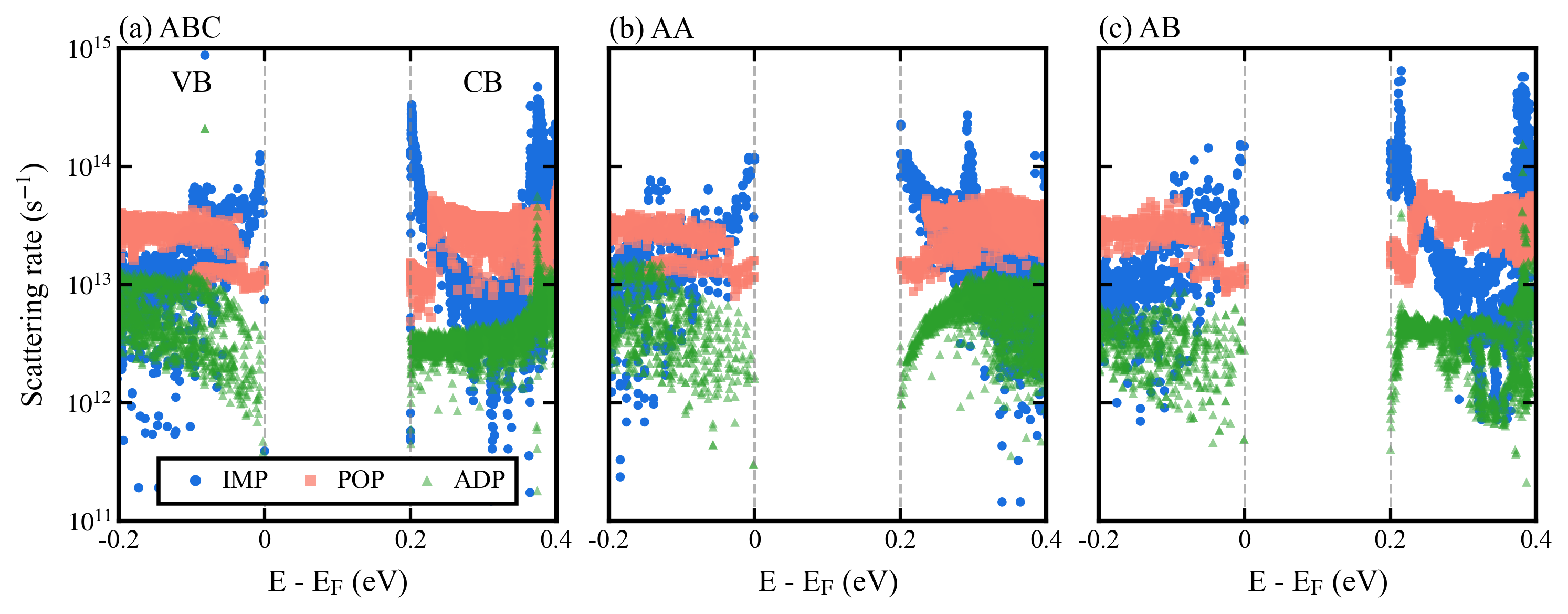}
		\caption{Electron-phonon scattering rates at 300~K, including acoustic deformation potential (ADP), polar optical phonon (POP), and ionized impurity (IMP) scattering, for (a) ABC, (b) AA, and (c) AB stacking configurations of Sc$_2$Si$_2$Te$_6$. For ease of comparison and visual clarity, the conduction band minimum (CBM) of all three configurations has been uniformly shifted to 0.2~eV.}
		\label{ss}
	\end{figure*}

	\subsection{Electronic transport properties.}
    Fig.~\ref{p} shows the temperature- and hole-concentration ($n_\mathrm{h}$)-dependent $\sigma$, $S$, and PF of Sc$_2$Si$_2$Te$_6$ for the three stacking sequences. Along both the in-plane and out-of-plane directions, $\sigma$ decreases with increasing temperature and increases with increasing $n_\mathrm{h}$, as commonly observed in semiconductors. In addition, $\sigma^{\perp}$ is substantially larger than $\sigma^{\parallel}$, consistent with the preceding analysis of the band effective mass $m^*_{\mathrm{b}}$. In contrast, $S$ exhibits the opposite trend to that of $\sigma$: it increases with increasing temperature and decreases with increasing $n_\mathrm{h}$. Moreover, $S^{\parallel}$ is slightly smaller than $S^{\perp}$. The opposite carrier-concentration dependences of $\sigma$ and $S$ give rise to an inverted-parabolic dependence of PF on $n_\mathrm{h}$, with a clear maximum at an optimal carrier concentration of $2\times10^{20}$~cm$^{-3}$. Consequently, for all stacking sequences, the maximum PF (PF$_\mathrm{max}$) along the in-plane direction is markedly larger than that along the out-of-plane direction, as shown in Table~\ref{pfmax}. For the three stacking sequences, the PF$_\mathrm{max}$ values occur at $n_\mathrm{h}$ $\approx$ 2 $\times$ 10$^{20}$~$\mathrm{cm}^{-3}$. At this concentration, the AB stacking exhibits the highest $\sigma$ (75725~S/m), whereas the AA stacking shows the lowest $\sigma$ (65709~S/m). At the same time, the $S$ value of AA is smaller than those of AB and ABC. Consequently, the AB and ABC stacking sequences possess much larger PF$_\mathrm{max}$ values than AA.

    Fig.~\ref{n} shows the temperature- and electron-concentration ($n_\mathrm{e}$)-dependent $\sigma$, $S$, and PF of the three Sc$_2$Si$_2$Te$_6$ stacking sequences. $\sigma$ exhibits pronounced anisotropy, with $\sigma^{\perp}$ being much larger than $\sigma^{\parallel}$, owing to the high energy of the conduction bands near the Fermi level along the out-of-plane direction. As a result, PF$^{\parallel}$ is close to 0.1~$\mu$W\,cm$^{-1}$\,K$^{-2}$. At 300~K and $n_\mathrm{e}$ = 10$^{20}$~cm$^{-3}$, corresponding to a Fermi level near the CBM, the $S$ values of the ABC, AA, and AB stacking structures are -220.7, -196.6, and -241.0~$\mu$VK$^{-1}$, respectively. This trend is consistent with the band degeneracy discussed above. A higher band degeneracy leads to a larger density of states (DOS) and larger $S$, as confirmed by the integrated density of states (IDOS) at the CBM, which are 37.7, 33.4, and 72.8 states/eV for the ABC, AA, and AB stacking sequences, respectively.

    Under $n$-type doping, as shown in Table~\ref{pfmax}, the PF$_\mathrm{max}$ values of all three structures occur at an electron concentration of 1 $\times$ 10$^{21}$~$\mathrm{cm}^{-3}$. At this concentration, the ABC stacking exhibits the highest $\sigma$ ($3.5 \times 10^{5}$~S/m), whereas the AA stacking has the largest magnitude of the $S$ ($-152$~$\mu$VK$^{-1}$). Nevertheless, the differences in $S$ among these stacking sequences are very small. Consequently, the ABC stacking exhibits the highest PF$_\mathrm{max}$ of 78.5~$\mu$Wcm$^{-1}$K$^{-2}$ at 800~K. The PF$_\mathrm{max}$ values of the other two structures are 76.1~$\mu$Wcm$^{-1}$K$^{-2}$ for AB and 69.9~$\mu$Wcm$^{-1}$K$^{-2}$ for AA, indicating relatively small differences among the stacking sequences. For a given structure, comparison of the in-plane transport under $p$- and $n$-type doping shows that PF$_\mathrm{max}$ under $n$-type doping is approximately three times larger than that under $p$-type doping. This enhancement arises from the high band degeneracy, which effectively balances $\sigma$ and $S$, allowing a relatively large $|S|$ of about 150~$\mu$V/K to be maintained even at a high carrier concentration of 1 $\times$ 10$^{21}$~$\mathrm{cm}^{-3}$. In summary, the largest PF$_\mathrm{max}$ values for all three structures are obtained along the out-of-plane direction under $n$-type doping. Moreover, the stacking sequence has only a minor influence on $S$, $\sigma$, and PF under heavily doped conditions, where the effect of band-edge degeneracy on the transport properties is reduced.

    As shown in Fig.~\ref{ss}, all three stacking sequences exhibit weak polar optical phonon (POP) scattering, which is the dominant scattering mechanism in ionic compounds. This weak POP scattering arises from the small ionic contributions to the dielectric response, $\varepsilon^{0}-\varepsilon^{\infty}$, particularly along the out-of-plane direction, as listed in Table~\ref{1111}. Therefore, the strongest scattering mechanism for electrons in these three stacking sequences is ionized-impurity (IMP) scattering, which is much stronger than that in Zintl compounds~\cite{doi:10.1021/acs.inorgchem.5c00031}. In the energy regions near the VBM and CBM, the ABC and AA sequences show similar IMP scattering rates, whereas the AB sequence exhibits slightly larger IMP scattering rates.

	\subsection{Phonon dispersion and lattice thermal conductivity.}
\begin{figure*}[!ht]
		\centering
		\includegraphics[width=1.0\linewidth]{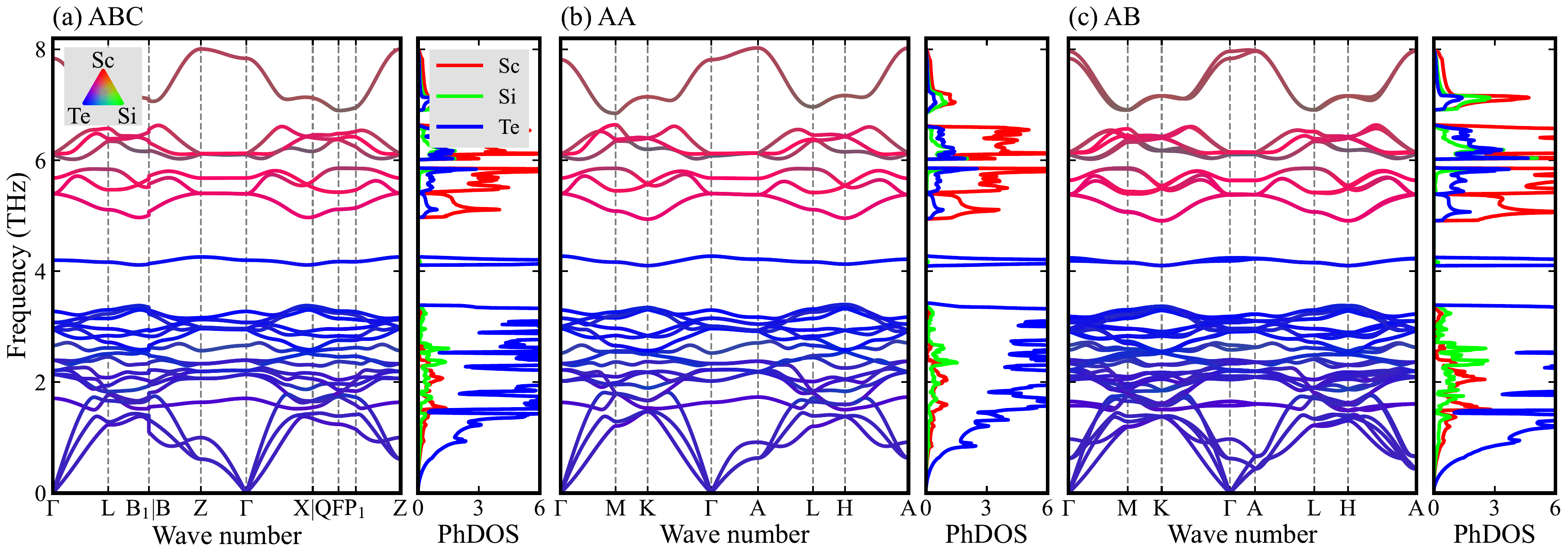}
		\caption{Phonon dispersion relations and phonon density of states (PhDOS) of Sc$_2$Si$_2$Te$_6$ at 300~K for (a) ABC, (b) AA, and (c) AB stacking sequences. Only the range of 0--8.2~THz is shown here, the full phonon dispersion is presented in Fig.~\textcolor{red}{S2}.}
		\label{phonon}
	\end{figure*}

	The phonon dispersions and phonon density of states (PhDOS) of Sc$_2$Si$_2$Te$_6$ with three different stacking structures at 300~K, calculated using the PBEsol functional, are shown in Fig.~\ref{phonon}. The primitive unit cells of the AA and ABC stacking sequences each contain 10 atoms, giving rise to 30 phonon branches, including 3 acoustic and 27 optical modes. In contrast, the primitive unit cell of the AB stacking contains 20 atoms, resulting in 60 phonon branches, including 3 acoustic and 57 optical modes. In Fig.~\ref{phonon}, the atomic contributions of Sc, Si, and Te to the phonon bands are color coded in red, green, and blue, respectively. Because the primitive cell sizes of the AA and ABC stacking sequences are identical and the vdW interaction is weak, their phonon dispersion are very similar. Specifically, the lowest optical phonon mode at the $\Gamma$ point is 1.63~THz for AA and 1.71~THz for ABC, while the highest acoustic phonon mode along the out-of-plane direction, i.e., $\Gamma$--A for AA and $\Gamma$--Z for ABC, is 0.92~THz for AA and 1.00~THz for ABC. By contrast, the AB stacking structure requires a doubled unit cell relative to AA, and therefore the Brillouin zone of AB is folded by a factor of two with respect to those of AA and ABC. Consequently, the frequencies of the lowest optical phonon mode at the $\Gamma$ point and the longitudinal acoustic (LA) phonon branch at the A point are 0.62 and 0.65~THz, respectively, nearly half of the corresponding values in the AA and ABC stacking structures.
	
	Apart from this difference, the phonon dispersions of the three structures are very similar. The phonon frequency $\omega$ is proportional to the square root of the bonding strength $k$ and inversely proportional to the square root of the atomic mass $M$, approximately following $\omega$ $\propto$ $\sqrt{k/M}$. Consequently, the heavier Te atoms, with an atomic mass of 127.60~a.u., predominantly contribute to the low-$\omega$ phonon modes in the range of 0--4~THz. The lightest Si atoms, with an atomic mass of 28.09~a.u., mainly contribute to the high-$\omega$ range of 10--16~THz, which is not shown in Fig.~\ref{phonon}. The Sc atoms, with an atomic mass of 44.96~a.u., are primarily located in the intermediate-$\omega$ range of 5--8~THz. This trend is also clearly reflected in the PhDOS, as shown in Fig.~\ref{phonon}.

	\begin{figure*}[!ht]
		\centering
		\includegraphics[width=1.0\linewidth]{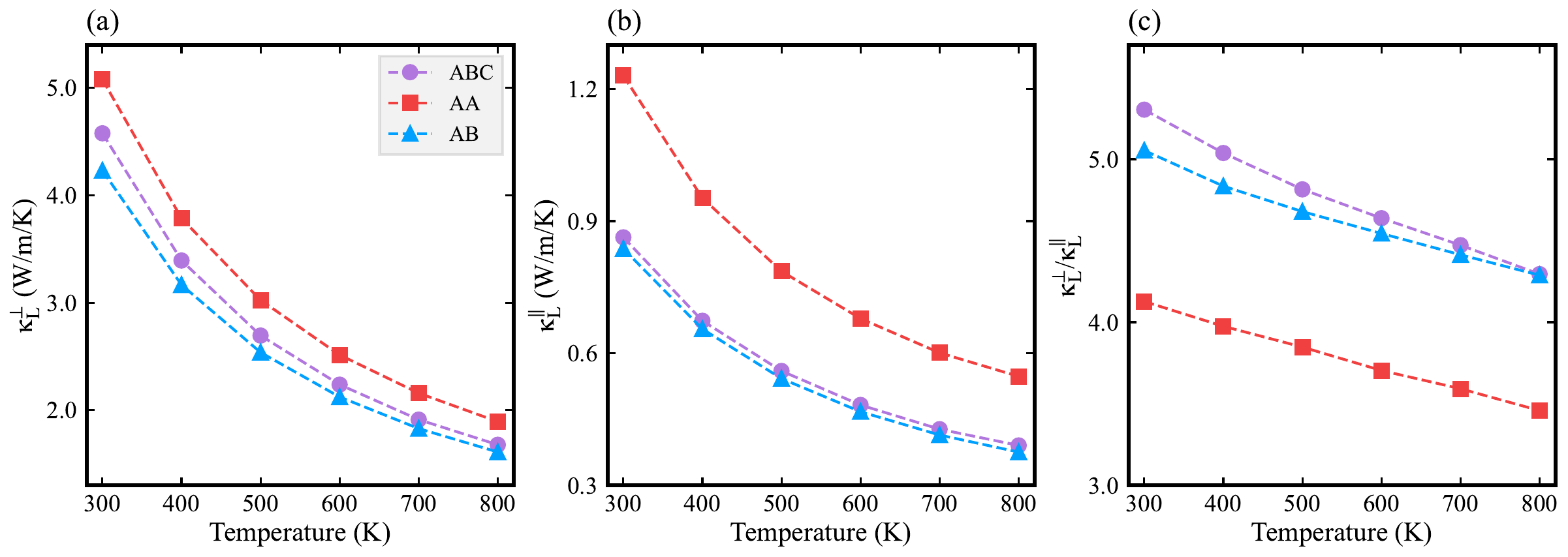}
		\caption{Temperature dependence of $\kappa_{\mathrm{L}}^{\perp}$ ($a$-axis), $\kappa_{\mathrm{L}}^{\parallel}$ ($c$-axis), and $\kappa_{\mathrm{L}}^{\perp}$/$\kappa_{\mathrm{L}}^{\parallel}$ ($a/c$) for Sc$_2$Si$_2$Te$_6$ with different stacking sequences.}
		\label{kappa}
	\end{figure*}

	\begin{figure*}[!ht]
		\centering
		\includegraphics[width=1.0\linewidth]{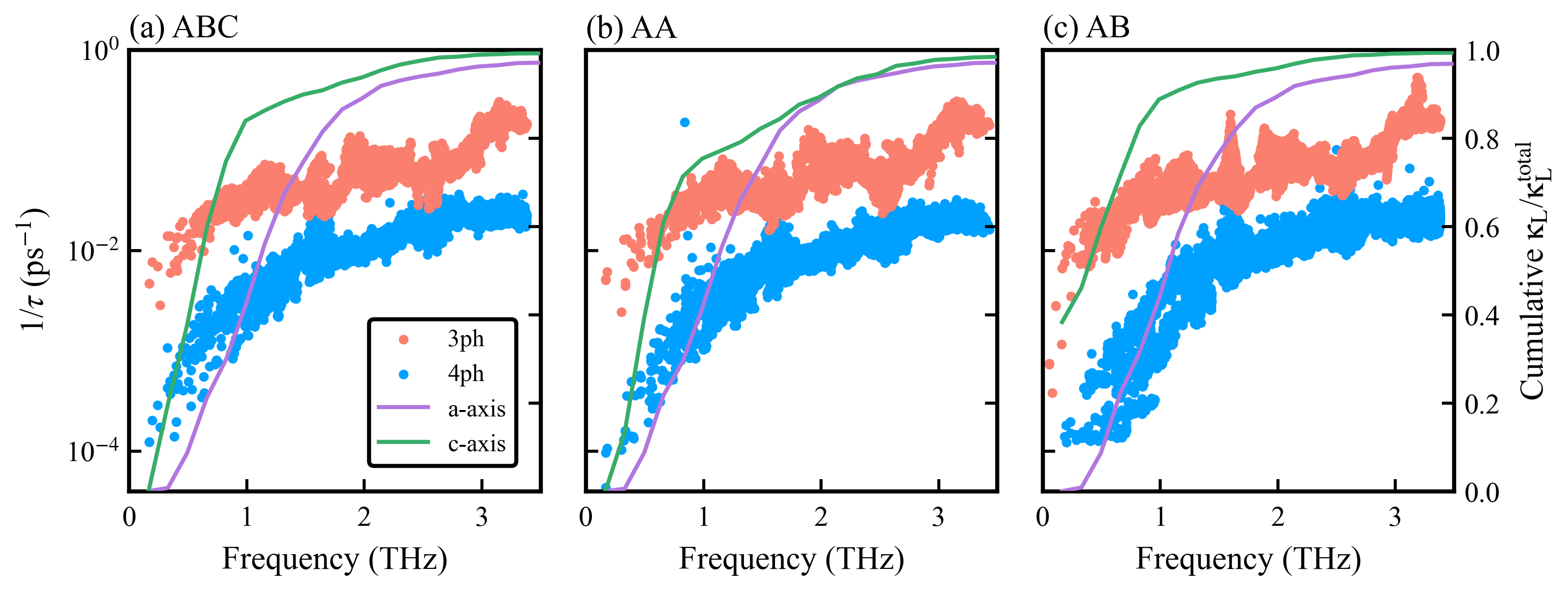}
		\caption{Three-phonon and four-phonon scattering rates, along with cumulative lattice thermal conductivity in the in-plane and out-of-plane directions, for Sc$_2$Si$_2$Te$_6$ at 300~K with (a) ABC, (b) AA, and (c) AB stacking sequences.}
		\label{phonon_ss}
	\end{figure*}

	Using the temperature-dependent renormalized second-order force constants, and including contributions from three-phonon (3ph) and four-phonon (4ph) scattering processes as well as coherent phonons, we calculated $\kappa_{\mathrm{L}}$ of the three Sc$_2$Si$_2$Te$_6$ stacking sequences, as shown in Fig.~\ref{kappa}. First, $\kappa_{\mathrm{L}}$ decreases with increasing temperature for all three structures along both the in-plane ($\kappa_{\mathrm{L}}^{\perp}$) and out-of-plane ($\kappa_{\mathrm{L}}^{\parallel}$) directions. Specifically, the temperature dependence of $\kappa_{\mathrm{L}}^{\perp}$ for AA, AB, and ABC follows $T^{-1.206}$, $T^{-1.213}$, and $T^{-1.189}$, respectively, whereas that of $\kappa_{\mathrm{L}}^{\parallel}$ follows $T^{-1.062}$, $T^{-1.085}$, and $T^{-1.074}$, respectively. Thus, $\kappa_{\mathrm{L}}^{\perp}$ exhibits a stronger temperature dependence than $\kappa_{\mathrm{L}}^{\parallel}$. Second, at all temperatures, both $\kappa_{\mathrm{L}}^{\perp}$ and $\kappa_{\mathrm{L}}^{\parallel}$ follow the order AA $>$ ABC $>$ AB. The AB stacking exhibits the lowest $\kappa_{\mathrm{L}}$, with $\kappa_{\mathrm{L}}^{\perp}$ = 4.23~Wm$^{-1}$~K$^{-1}$ and $\kappa_{\mathrm{L}}^{\parallel}$ = 0.84~Wm$^{-1}$~K$^{-1}$ at room temperature, which further decrease to 1.61 and 0.37~Wm$^{-1}$K$^{-1}$ at 800~K, respectively. Nevertheless, the differences among the three stacking sequences are relatively small along both directions, owing to the weak influence of the vdW interactions on the stacking-dependent thermal transport.
	
	All three stacking sequences exhibit pronounced anisotropy in $\kappa_{\mathrm{L}}$, as characterized by the ratio $\kappa_{\mathrm{L}}^{\perp}/\kappa_{\mathrm{L}}^{\parallel}$. Among them, AA shows the smallest anisotropy, whereas ABC exhibits the largest. The larger $\kappa_{\mathrm{L}}^{\perp}$ compared with $\kappa_{\mathrm{L}}^{\parallel}$ is mainly attributed to the larger phonon group velocities within the layers, as shown in Fig.~\textcolor{red}{S3}. Notably, a previous theoretical study on ABC reported a $\kappa_{\mathrm{L}}^{\perp}/\kappa_{\mathrm{L}}^{\parallel}$ ratio of 2.24 at 300~K~\cite{Sc2Si2Te6}, which is much smaller than our calculated value of 5.30. This discrepancy mainly arises from the choice of exchange-correlation functional and dispersion correction. In the previous calculation~\cite{Sc2Si2Te6}, the D3 dispersion correction was applied to the PBEsol functional. However, the D3 correction was not originally designed for PBEsol~\cite{10.1063/1.3382344}. Moreover, PBEsol generally performs well for vdW systems~\cite{PhysRevB.89.075409}, and the lattice constants calculated using PBEsol agree well with the experimental values, as shown in Table~\ref{1111}. By contrast, the interlayer vdW interaction is likely overestimated by the PBEsol+D3 method, as evidenced by the 5\% underestimation of the lattice constant $c$ compared with the experimental value~\cite{https://doi.org/10.1002/zaac.202200234}. The enhanced interlayer interaction leads to an overestimated $\kappa_{\mathrm{L}}^{\parallel}$, thereby yielding a smaller $\kappa_{\mathrm{L}}^{\perp}/\kappa_{\mathrm{L}}^{\parallel}$ ratio.
	
	Fig.~\ref{phonon_ss} presents the 3ph and 4ph scattering rates, together with the cumulative $\kappa_{\mathrm{L}}$ as a function of phonon frequency, for the three Sc$_2$Si$_2$Te$_6$ structures. The cumulative $\kappa_{\mathrm{L}}$ shows that phonons with frequencies below 2~THz contribute approximately 90\% of the total $\kappa_{\mathrm{L}}$, highlighting the dominant role of acoustic and low-frequency optical phonon modes. In this frequency range, the 3ph scattering rates are significantly larger than the 4ph scattering rates and therefore dominate the phonon scattering processes that limit $\kappa_{\mathrm{L}}$. Among the three stacking sequences, AB exhibits the largest 3ph scattering rates, whereas AA shows the smallest. As shown in Fig.~\textcolor{red}{S4}, the 4ph contribution to $\kappa_{\mathrm{L}}$, defined as $(\kappa_{\mathrm{L}}^{\mathrm{3ph}}-\kappa_{\mathrm{L}}^{\mathrm{3,4ph}})/\kappa_{\mathrm{L}}^{\mathrm{3ph}}$, is larger for the ABC stacking sequence along the in-plane direction than for the AA and AB stacking sequences, owing to its slightly larger 4ph scattering rates. In addition, Fig.~\textcolor{red}{S3} shows that the phonon group velocity $\nu_\mathrm{g}$ of the AB stacking sequence is smaller than those of AA and ABC, especially along the out-of-plane direction. Consequently, AB exhibits the lowest $\kappa_{\mathrm{L}}$ among the three stacking sequences.

	\begin{figure*}[!ht]
		\centering
		\includegraphics[width=1.0\linewidth]{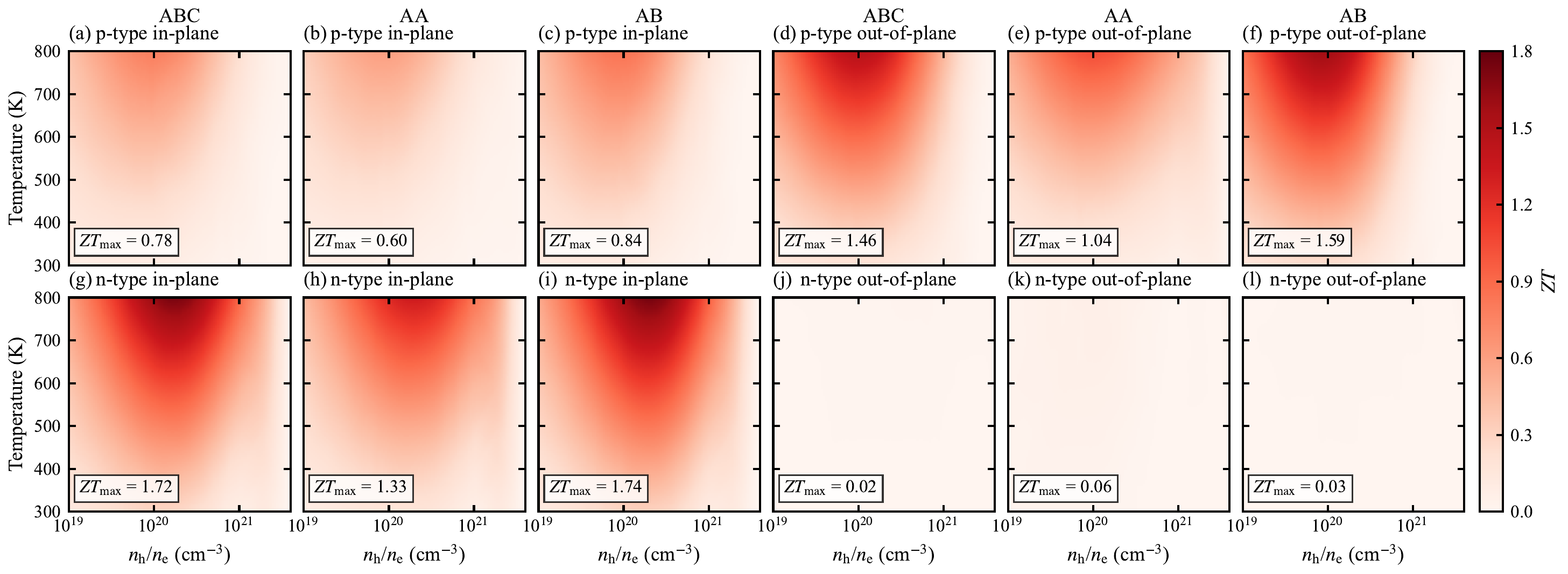}
		\caption{Thermoelectric figure of merit ($ZT$) of Sc$_2$Si$_2$Te$_6$ for three different stacking sequences: (a)--(c) in-plane $ZT$ under $p$-type doping, (d)--(f) out-of-plane $ZT$ under $p$-type doping, (g)--(i) in-plane $ZT$ under $n$-type doping, and (j)--(l) out-of-plane $ZT$ under $n$-type doping.}
		\label{zt}
	\end{figure*}

	\subsection{Figure of merit $ZT$.}	
	After obtaining the electronic and thermal transport properties, we evaluated the carrier-concentration and temperature dependence of the dimensionless figure of merit $ZT$ for the three stacking structures, as shown in Fig.~\ref{zt}. Since $ZT$ is proportional to temperature and inversely proportional to $\kappa_{\mathrm{L}}$, and since $\kappa_{\mathrm{L}}$ decreases with increasing temperature, the maximum $ZT$ ($ZT_\mathrm{max}$) for these structures always occurs at 800~K, i.e., the upper limit of the temperature range considered in this work. This temperature is still safely below the thermal stability limit of 1023~K~\cite{https://doi.org/10.1002/zaac.202200234}. In addition, $ZT$ exhibits the characteristic behavior of first increasing and then decreasing with increasing carrier concentration. For both $p$-type and $n$-type doping, $ZT_\mathrm{max}$ appears near $10^{20}$~cm$^{-3}$.
	
	For $p$-type doping, the $ZT$ values along the out-of-plane direction ($ZT^{\parallel}$) are significantly larger than those along the in-plane direction ($ZT^{\perp}$), primarily because $\kappa_{\mathrm{L}}^{\parallel}$ is much smaller, as shown in Fig.~\ref{kappa}, even though PF$^{\perp}$ exceeds PF$^{\parallel}$. Among the different stacking sequences, AB exhibits the highest $ZT_{\mathrm{max}}^{\parallel}$ (1.59), followed by ABC (1.46) and AA (1.04). Under $n$-type doping, the PF$^{\parallel}$ values of all three sequences are very small, as shown in Fig.~\ref{n}, owing to their low $\sigma^{\parallel}$; consequently, the corresponding $ZT^{\parallel}$ values are also close to zero. Along the in-plane direction, AB again yields the largest $ZT_\mathrm{max}^{\perp}$, with a peak value of 1.74, followed by ABC and AA, with $ZT_\mathrm{max}^{\perp}$ values of 1.72 and 1.33, respectively. Therefore, the thermoelectric performance of these three stacking sequences follows the same order under both $p$-type and $n$-type doping: AB $>$ ABC $>$ AA. This trend is consistent with the order of $\kappa_{\mathrm{L}}$ from smallest to largest, namely AB $<$ ABC $<$ AA. The underlying reason is that different stacking sequences have only a small effect on electronic transport under heavy doping, but they exert a relatively strong influence on $\kappa_{\mathrm{L}}^{\parallel}$.
	
	Note that a carrier concentration of $2\times10^{20}$~cm$^{-3}$ may be difficult to realize experimentally in many compounds. Therefore, we compare the electronic transport properties and $ZT$ values of the three stacking sequences at a carrier concentration of $5\times10^{19}$~cm$^{-3}$ and 800~K. For hole doping, the PF$^{\perp}$ (PF$^{\parallel}$) values of the ABC, AA, and AB stacking sequences are 17.58, 15.60, and 18.66~$\mu$W/cm/K$^{2}$ (8.54, 8.40, and 9.37~$\mu$W/cm/K$^{2}$), respectively, and the corresponding $ZT^{\perp}$ ($ZT^{\parallel}$) values are 0.72, 0.57, and 0.79 (1.35, 0.96, and 1.51). For electron doping, the PF$^{\perp}$ (PF$^{\parallel}$) values of the ABC, AA, and AB stacking sequences are 32.72, 27.17, and 29.17~$\mu$W/cm/K$^{2}$ (0.04, 0.37, and 0.03~$\mu$W/cm/K$^{2}$), respectively, and the corresponding $ZT^{\perp}$ ($ZT^{\parallel}$) values are 1.34, 1.00, and 1.26 (0.01, 0.05, and 0.01). These results indicate that the influence of the stacking sequence on the thermoelectric properties is more pronounced at lower doping concentrations than under heavily doped conditions. The AA stacking structure consistently exhibits the poorest thermoelectric performance among the three stacking sequences.

	\maketitle
	\section{Conclusions}
	In this work, we performed a systematic first-principles investigation of the thermodynamic stability and thermoelectric properties of Sc$_2$Si$_2$Te$_6$ with three high-symmetry stacking sequences, namely AA, AB, and ABC. The AB stacking sequence is nearly energetically degenerate with the experimentally reported ABC structure, whereas the AA stacking structure is only 7~meV/f.u. higher in energy than the ABC stacking structure. The maximum energy barrier for transitions among these stacking sequences is approximately 10~meV/atom, comparable to those of other vdW layered compounds such as MoS$_2$, in which multiple stacking polymorphs have been experimentally observed. This result is also consistent with the experimentally observed stacking faults in Sc$_2$Si$_2$Te$_6$~\cite{https://doi.org/10.1002/zaac.202200234}.
	
	Electronic structure calculations show that the valence band maximum of all three stacking sequences is located at the $\Gamma$ point, whereas the conduction band minimum is strongly stacking dependent. Specifically, the conduction band minimum of ABC lies at a point between two high-symmetry points, that of AA is located at the K point, and that of AB occurs at the K and H points with a negligible energy difference. Accordingly, the conduction-band degeneracies of ABC, AA, and AB are 12, 2, and 8, respectively. These electronic-structure characteristics have a clear influence on the electronic transport properties at doping levels for which the Fermi level is close to the conduction band minimum. For example, at a carrier concentration of $5 \times 10^{19}$~cm$^{-3}$ and 300~K, the in-plane power factors of the ABC, AA, and AB stacking sequences are 39.97, 26.09, and 34.74~$\mu$W~cm$^{-1}$~K$^{-2}$, respectively. Regardless of doping level, the in-plane power factors of all three stacking structures are consistently superior to their out-of-plane counterparts under both $p$-type and $n$-type doping, owing to the large hole effective mass and high conduction-band energy along the out-of-plane direction. However, at high doping levels, where large electrical conductivity is achieved, the stacking sequence has only a weak effect on the power factor. The maximum power factor is obtained for the ABC stacking, 78.5~$\mu$W~cm$^{-1}$~K$^{-2}$, followed by AB, 76.1~$\mu$W~cm$^{-1}$~K$^{-2}$, and AA, 69.9~$\mu$W~cm$^{-1}$~K$^{-2}$.
	
	Our calculations demonstrate that three-phonon scattering plays a more important role than four-phonon scattering in suppressing heat transport in these structures. Among the three stacking sequences, AB exhibits the lowest lattice thermal conductivity, owing to its large three-phonon scattering rates and low sound velocity. The stronger four-phonon scattering in the ABC structure results in a lower lattice thermal conductivity than that of the AA stacking structure. All three stacking sequences exhibit pronounced anisotropy, with lower lattice thermal conductivities along the out-of-plane direction, due to the weak van der Waals interactions between layers. Consequently, the maximum $ZT$ values of all three structures occur along the in-plane direction under $n$-type doping, mainly due to their favorable in-plane electronic transport properties. The $ZT$ values decrease in the order 1.74 for AB, 1.72 for ABC, and 1.33 for AA. In addition, under $p$-type doping, the out-of-plane direction also exhibits relatively large $ZT$ values, namely 1.59 for AB, 1.46 for ABC, and 1.04 for AA, primarily resulting from the extremely low lattice thermal conductivities along the out-of-plane direction. Our study reveals that different stacking configurations in Sc$_2$Si$_2$Te$_6$ have a non-negligible effect on thermoelectric transport properties, and a pronounced reduction in $ZT$ is expected when the AA stacking sequence is present.

	\section{Acknowledgments}
	The authors acknowledge the support of the National Science Foundation of China (Grant No. 12374024) and supported by State Key Laboratory for Advanced Metals and Materials, Grant No. 2025-Z15.

	\bibliography{ref}
	
\end{document}